\documentclass[a4paper]{jpconf}
\usepackage{graphicx}
\usepackage[svgnames]{xcolor}
\usepackage{hyperref}
\usepackage{xspace}
\usepackage{relsize}
\usepackage[frozencache,cachedir=minted-cache]{minted}
\definecolor{darkCol}{HTML}{222427}

\newcommand{\podio}{\textrm{podio}\xspace}
\newcommand{\edmhep}{\textrm{EDM4hep}\xspace}
\newcommand{\Rplus}{\protect\hspace{-.1em}\protect\raisebox{.35ex}{\smaller{\smaller\textbf{+}}}}
\newcommand{\cpp}{\mbox{C\Rplus\Rplus}\xspace}

\begin{document}

\title{Of Frames and schema evolution - The newest features of podio}
\author{Placido Fernandez Declara$^{1}$,
  Frank Gaede$^{2}$,
  Gerardo Ganis$^{1}$,
  Benedikt Hegner$^{1}$*,
  Clement Helsens$^{3}$,
  Thomas Madlener$^{2}$*,
  Andre Sailer$^{1}$,
  Graeme A Stewart$^{1}$ and
  Valentin Volkl$^{1}$
}

\address{$^{1}$CERN, Switzerland}
\address{$^{2}$Deutsches Elektronen-Synchrotron DESY, Germany}
\address{$^{3}$KIT, Germany}

\ead{benedikt.hegner@cern.ch, thomas.madlener@desy.de}

\begin{abstract}
  The podio event data model (EDM) toolkit provides an easy way to generate a
  performant implementation of an EDM from a high level description in yaml
  format. We present the most recent developments in podio, most importantly the
  inclusion of a schema evolution mechanism for generated EDMs as well as the
 ``Frame'', a thread safe, generalized event data container. For the former we
  discuss some of the technical aspects in relation with supporting different
  I/O backends and leveraging potentially existing schema evolution mechanisms
  provided by them. Regarding the Frame we introduce the basic concept and
  highlight some of the functionality as well as important aspects of its
  implementation. The usage of podio for generating different EDMs for future
  collider projects (most importantly EDM4hep, the common EDM for the Key4hep
  project) has inspired new features. We present some of those smaller new features and end with a brief overview on current developments towards a first
  stable version as well as an outlook on future developments beyond that.
\end{abstract}

\section{Introduction}\label{sec:introduction}

\podio is an event data model (EDM) toolkit that allows one to generate thread safe
and efficient \cpp code from a high level description in yaml format. It favors
composition over inheritance and uses plain-old-data (POD) types wherever
possible. EDMs generated by \podio feature a User Layer consisting of thin handles
offering value semantics, which is powered by two more layers that manage
resources and relations between objects. This layered approach also allows one to
support multiple I/O backends, where ROOT~\cite{Brun:1997pa} is the default and
an alternative based on \emph{Simple Input/Output} (SIO)~\cite{sio:github} is
also available. For more details about how \podio generates EDMs we refer
to previous publications~\cite{podio,podio_2020,Gaede:2021izq}. One of the main use cases for \podio
at the moment is the generation of \edmhep~\cite{Gaede:2021izq,Gaede:2022leb},
the EDM of the Key4hep
project~\cite{FernandezDeclara:2022voh,Fang:2023mwt,key4hep_2021,Volkl:20225e}.

In these proceedings we discuss a few recent developments in the \podio toolkit.
One is the introduction of the \emph{Frame} concept, where we introduce the
basic ideas and some implementation details in Sec.~\ref{sec:frame}. Another
important recent development is related to schema evolution of generated EDMs.
This has been a longstanding missing feature of \podio, and we discuss the plans
and currently existing functionality in Sec.~\ref{sec:schema_evolution}. We also
give brief descriptions of other, smaller recent developments in
Sec.~\ref{sec:other-devel} before we conclude with a brief outlook.


\section{The Frame concept and design}\label{sec:frame}

One of the shortcomings of \podio so far was its usability in multi-threaded
contexts. The \emph{EventStore} that has been shipped with \podio was never
designed to support this, and has far outlived its original purpose of an
example implementation for a transient event store. The \emph{Frame} concept has
been developed to address these short comings and to provide a more production
ready way of accessing and operating on data. The main design goals for the
Frame concept are
\begin{itemize}
  \item Serve as a container that aggregates all relevant data
  \item Define an \emph{interval of validity} or category for the the contained
        data
  \item Easy to use and thread-safe interface for accessing and storing data
  \item Clearly defined ownership and mutability of data that is also reflected
        in the interface, while still supporting value semantics
  \item Separation of reading data and the necessary processing in order to,
        e.g. do schema evolution or establish inter-object relations
\end{itemize}

By keeping the Frame concept rather general and not prescribing any ad-hoc
definition of its category or interval of validity, we aim at a more general
concept than its potential main use case as a HEP event data container. Since
the contained data and the user define the category of each individual Frame it
should also be usable by experiments that have no clear notion of an ``event'',
but rather operate with, e.g., readout time frames.

\begin{figure}[h]
  \centering
  \includegraphics[width=0.4\linewidth]{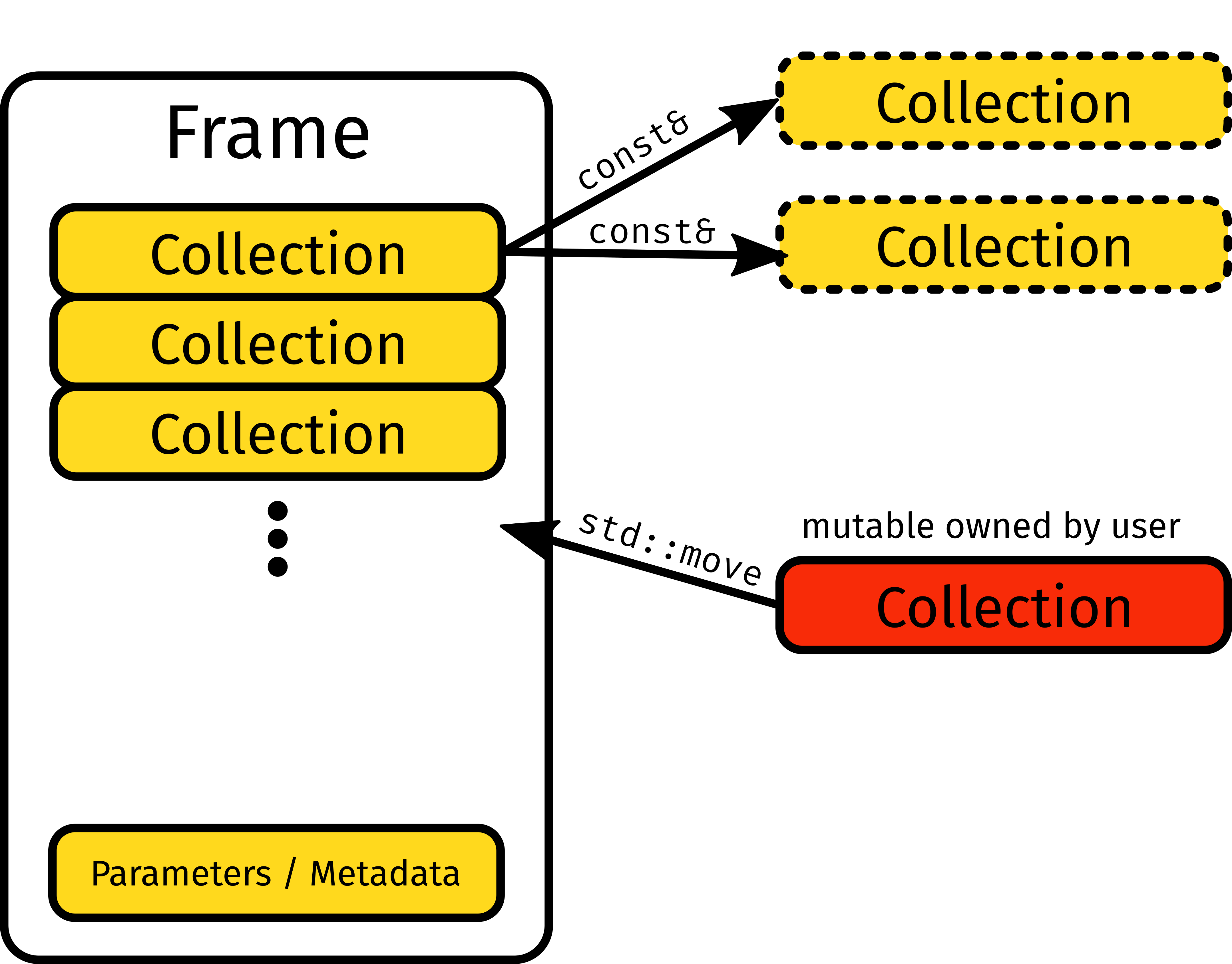}
  \caption{Schematic overview of the Frame concept and the ownership and mutability of collection data.}
  \label{fig:frame-concept}
\end{figure}

The design choices for achieving a thread-safe interface and for clearly
defining ownership of the contained data are closely related, and the former
more or less necessitates the latter. The easiest way to make concurrent access
to collections stored in a Frame possible is to ensure that these accesses are
read-only and so, by definition, do not mutate stored data. In order to ensure
that users cannot keep a mutable reference to collections they put into a Frame
they have to explicitly relinquish ownership by \emph{moving} collections into
the Frame, as shown in Fig.~\ref{fig:frame-concept}. Although \cpp does not have
the notion of a destructive move it is possible to enforce this ownership
transfer at compile time. One of the drawbacks of this approach is that it is
still possible to use \emph{moved-from} collections. Even though this does not
touch the thread-safety properties of the Frame, the results might be confusing
for non-experienced \cpp users. Here we have to rely on tooling, extensive
documentation and training of users to not use moved-from objects in \cpp.

For the implementation of the Frame class we decided to use type erasure. This
allows us to have a move-only type that still has value semantics. It is also
required by the goal of clearly decoupling I/O from the Frame interface, as it
allows us to make Frames constructible from almost arbitrary \emph{Frame data}.
With this possibility each I/O backend can define its own implementation of such
data that optimally uses its capabilities, as long as the provided
implementation fulfills a rather minimal interface to actually access the POD
buffers from which a collection can be constructed.

The basic assumptions for I/O in the Frame concept are that each file that is
read from or written to is only operated on from a single thread. This
assumption is exploited inside the Frame by assuming that no other thread has
direct access to the Frame data from which it has been constructed, greatly
simplifying the necessary synchronization. Another aspect that contributed to
this decision is the much simpler implementation of basic readers and writers,
which now exist for both available backends, ROOT and SIO. Here we see \podio as
a provider of basic building blocks that can be combined by downstream users in
more complex scenarios, e.g. having one thread producing Frame data with a
reader and several threads consuming these to construct Frames from.

All the basic functionality for the Frame concept is in place, and the main work
for the near future is to replace all usages of the EventStore with Frame based
I/O in the Key4hep stack. Another possibility in the current design is the
implementation of policies that change the Frame behavior without changing its
interface.


\section{Schema evolution in podio}\label{sec:schema_evolution}

Since detector technologies and reconstruction algorithms are ever evolving, the
datatypes that are used to store all the necessary data have to be able to
evolve as well. This evolution has to happen in a way that grants access to data
that has been written with previous versions. Hence, schema evolution is a
crucial feature for \podio generated EDMs.

A general schema evolution mechanism is highly non-trivial to implement since
the problem space that needs to be covered is almost unbounded. Rather than
trying to tackle this challenge, \podio focuses on providing the necessary hooks
to do schema evolution, and to be able to leverage existing schema evolution
capabilities of existing I/O backends. With the hooks in place, the actually
supported schema evolutions will be implemented as the use cases arise in the
communities that use \podio. For these implementations the goal is to have
automatic code generation for as many cases as possible, but to still have the
possibility for user defined evolution functions where necessary.

As a first piece of the solution we have implemented a tool that reads the high
level yaml format definition of the same EDM in two different schema versions to
produce a list of differences between the two. An example result of this
tool is shown in Listing~\ref{lst:schema_diff}. As can
be seen from the result the tool is able to categorize the detected schema
changes into supported and (currently) unsupported schema changes. At this point
in time this categorization targets the ROOT backend and its builtin schema
evolution capabilities. An important aspect of this pre-processing step of
running the comparison tool is that it allows one to check beforehand whether a
given schema change is currently supported. Without this check it would be
easily possible to lose access to already written data.
\begin{listing}
  \begin{minted}[
    fontsize=\footnotesize,
    %bgcolor=darkCol
]{console}
Comparing datamodel versions v2 and v1

Found 4 schema changes:
 - 'ToBeDroppedStruct' has been dropped
 - 'ex2::NamespaceStruct' has an addded member 'y'
 - 'ex2::NamespaceStruct' has a dropped member 'y_old'
 - 'ExampleStruct.x' changed type from int to double

Warnings:
 - Definition 'ex2::NamespaceStruct' has a potential member rename 'y_old' -> 'y' of type 'int'.

ERRORS:
 - Forbidden schema change in 'ex2::NamespaceStruct' for 'x' from 'std::array<int, 2>' to 'int'
  \end{minted}
\caption{Example result of the analysis of two datamodel definitions with different schema versions.}
\label{lst:schema_diff}
\end{listing}

\begin{figure}[h]
  \centering
  \includegraphics[width=0.5\linewidth]{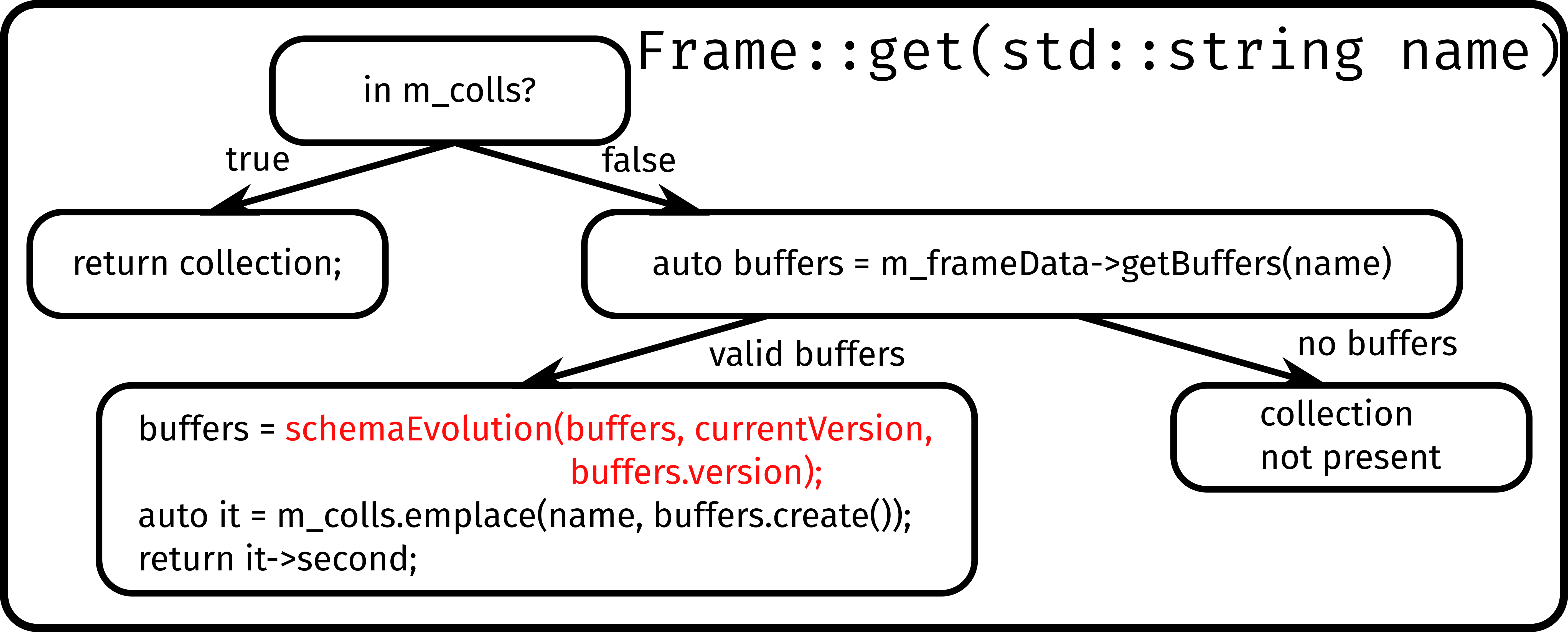}
  \caption{Conceptual implementation of getting a collection from a Frame,
    including reading the POD buffers from the internal Frame data and a
    potential evolution of these buffers (red) before the construction of a
    collection.}
  \label{fig:frame_get_concept}
\end{figure}

The next step for supporting schema evolution also for other backends than ROOT
is to put the necessary schema evolution hooks into place.
Fig.~\ref{fig:frame_get_concept} shows where we plan to place these hooks within
the Frame method for retrieving a collection. In this concept the schema
evolution hooks are placed at the earliest point in time where they can possibly
be; immediately after getting the POD buffers from the internal Frame data. In
case an I/O backend has already performed schema evolution on these buffers, the
hook essentially becomes a no-op function call, otherwise the hook will make
sure to call the correct evolution function, depending on the requested
collection type and involved schema versions. As a consequence of this placement
of the schema evolution hook, users will only ever see the datatypes in their
latest version.

The last step towards full schema evolution support in \podio generated EDMs is
the automatic code generation from the list of differences between two schema
versions. Here we have the basics in place with the comparison tool, and plan to
implement a prototype shortly.


\section{Other recent developments}\label{sec:other-devel}

Apart from these bigger developments that have just been described, there also
have been some smaller improvements, that were triggered by different
communities that are using \podio.

\paragraph{JSON output}

The Phoenix event display framework~\cite{phoenix_eventdisplay} expects the
event data that should be displayed to be in JSON format. In order to support
this use case the possibility to dump collections of \podio generated EDMs to
JSON has recently been implemented using the \emph{nlohmann/json}
library~\cite{nlohmann_json_2022}. We would like to point out that there are no
plans to also support reading from JSON format in \podio, and this is considered
purely an output format.

\paragraph{Datamodel extensions}

In many future collider contexts the contents and necessary relations for
datatypes that should represent the measured data of novel detector concepts are
not yet established and some prototyping is necessary. To facilitate this
prototyping phase we have recently implemented the possibility to extend
existing datamodel definitions with new datatypes. By specifying an
\emph{upstream datamodel}, the datatypes and components defined in this upstream model become
available for use in the datamodel extension. This is currently used by several
detector concepts that extend the existing \edmhep definition in order to
develop novel reconstruction algorithms. By installing the high level yaml
description file alongside the build artifacts \edmhep endorses this approach
for prototyping. As we have already done so in other places~\cite{Gaede:2022leb}
we would like to point out again, that this feature should be used solely for
prototyping.

Another accomplishment that is made possible by this extension mechanism is the
usage of \edmhep types in EDM4eic~\cite{edm4eic:github}, the datamodel used by the Electron-Ion
Collider (EIC) community. The previous practice of re-defining the same
datatypes again has now been replaced by the extension mechanism making this
dependency much clearer. However, also in this use case the long term goal is to
eventually converge on one set of datatypes defined in \edmhep.


\section{Conclusion \& Outlook}\label{sec:conlusion}

The \podio EDM toolkit has received crucial new developments recently; The Frame
concept should allow for better usability in multi-threaded contexts, but also
for a clearer separation of concerns with respect to I/O operations. Another
longstanding issue is addressed by the implementation of schema evolution
capabilities for generated EDMs. Here we have laid important ground work and
have some clear steps ahead to get this feature fully functional. Additionally,
there have been some smaller developments, mostly inspired by use cases from the
communities using \podio.

The next steps for \podio are to finish schema evolution capabilities, as well
as the integration of Frame based I/O functionality into more Key4hep
components. We plan to release a first stable version of \podio once all the
necessary hooks for schema evolution are in place.

\ack{This work benefited from support by the CERN Strategic R\&D Programme on
Technologies for Future Experiments (\url{https://cds.cern.ch/record/2649646/},
CERN-OPEN-2018-006) and has received funding from the European Union's Horizon
2020 Research and Innovation programme under grant agreement No 101004761.}

\section*{References}
\bibliographystyle{iopart-num.bst}
\bibliography{bibliography}

\end{document}